\documentstyle[aps,epsfig,prb]{revtex}                          %twocolumn
\begin{document}
\twocolumn[\hsize\textwidth\columnwidth\hsize\csname     
%twocolumn
@twocolumnfalse\endcsname                               %twocolumn 
\draft

\title{Phase-Sensitive Tetracrystal Pairing-Symmetry Measurements and
Broken Time-Reversal Symmetry States of High $T_c$ Superconductors}

\author{M. B. Walker}
\address{Department of Physics, 
University of Toronto,
Toronto, Ont. M5S 1A7 }
\date{\today }
\maketitle         

\widetext                                       %
\begin{abstract}
  A detailed analysis of the symmetric tetracrystal geometry used in
phase-sensitive pairing symmetry experiments on high T$_c$
superconductors is carried out for both bulk and surface time-reversal
symmetry-breaking states, such as the $d+id^\prime$ and $d+is$ states.
The results depend critically on the substrate geometry.  In the
general case, for the bulk $d+id^\prime$ (or $d+is$) state, the
measured flux quantization should in general not be too different from
that obtained in the pure $d$-wave case, provided $|d^\prime| \ll
|d|$ (or $|s| \ll |d|$).  However, in one particular high symmetry geometry, the $d+id^\prime$ state gives results that
allow it to be distinguished from the pure $d$ and the $d + is$
states. Results are also given for the cases where surface $d+is$ or
$d+id^\prime$ states occur at a {\rm [110]} surface of a bulk $d$-wave
superconductor.  Remarkably, in the highest symmetry geometry, a number of the broken time-reversal symmetry states discussed above give flux quantization conditions usually 
associated with states not having broken time-reversal symmetry.
\end{abstract}


\vfill          %twocolumn
\narrowtext                     %

\vskip2pc]      %twocolumn

%\newpage

This article describes the use of the symmetric tetracrystal
geometry in phase-sensitive pairing-symmetry experiments designed to investigate broken time-reversal symmetry superconducting states.  The advantages of the symmetric tetracrystal geometry for such experiments will be outlined, and  results will be given for a number of specific cases.

All of the hole-doped high temperature superconductors (such as
YBa$_2$Cu$_3$O$_{6+x}$) measured up to the present in phase sensitive
pairing symmetry experiments have exhibited the behavior expected of a superconducting state with $x^2 -
y^2$ symmetry (see Ref.\ \protect\onlinecite{tsu00}).  However, interesting
questions related to the symmetry of the superconducting state of the
high T$_c$ superconductors still remain, and a number of
investigations of possible deviations from a pure $x^2 - y^2$ symmetry
have been and are being undertaken. For example, it has been suggested
\protect\cite{bal98} that
at low temperatures, an interaction with magnetic impurities can cause the
$d$-wave state to become unstable with respect to the formation of a 
$d+id^\prime$ state. (Here $d$ is used to denote a state of
$x^2 - y^2$ symmetry, and $d'$ to denote a state of $xy$ symmetry.) Also,
the anyon mechanism for high T$_c$ superconductivity is known 
\protect\cite{rok93} to give
superconducting states of $d+id^\prime$ symmetry.  Attempts to interpret
\protect\cite{bal98,lau98}
anomalies observed \protect\cite{mov98,kri98} in the thermal conductivity of certain bismuth superconductors in terms of $d+id^\prime$ superconductivity
has further heightened interest in broken time-reversal symmetry states.
Bulk $d+is$ superconductivity has also been noted as a possibility
\protect\cite{li93}.
[It is known that because the $d+id^\prime$ state (and
the $d+is$ state) involves order parameters ($d$ and $d^\prime$) with
two different symmetries, the phase transition to a superconducting $d+id^\prime$ state can
not occur at a single continuous phase transition.] 
Also, the possible existence of surface $d+is$ or $d+id^\prime$ states at a
[110] surface of a bulk $d$-wave superconductor has been explored
theoretically in Refs.\ \protect\onlinecite{mat95,kas95,fog97,zhu98,zhu99},
particularly with reference to the possible
splitting of the zero bias conductance peak observed in NIS (normal
to insulator to superconductor) tunneling experiments.  Experimental
evidence for such a splitting has been found, at low temperatures, in
Ref. \protect\onlinecite{cov97}.

Very recently, evidence \protect\cite{tsu00b} of the half integral
flux quantum effect in tricrystal experiments on the electron-doped
superconductors Nd$_{2-x}$Ce$_x$CuO$_{4}$ and Pr$_{2-x}$Ce$_x$CuO$_4$
has been presented indicating that the superconducting state of these
materials also has $d$-wave symmetry.   This result should be contrasted 
with one other recent experimental result, namely a thermal
conductivity study \protect\cite{tai00} on Pr$_{2-x}$Ce$_x$CuO$_4$
indicating that there are no propagating low-energy
quasiparticle excitations in this material.  These results (in which a
material displays both a gapless $d$-wave state and the absence of
propagating low energy excitations) are extremely unusual, and while
perhaps not necessarily contradictory, require further study if they
are to be understood. One possibility is that the absence of
propagating low energy states implied by the thermal conductivity
measurements could be accounted for if the low energy quasiparticle
excitations were localized by disorder in Pr$_{2-x}$Ce$_x$CuO$_4$, as
pointed out in Ref.\ \protect\onlinecite{tai00} (see also Ref.\ 
\protect\onlinecite{hus00}).  

A question that might be raised about these results is to what extent the observations of a half integral flux quantum in a tricrystal experiment could, rather than implying a pure $d$-wave state, be consistent with a $d + id'$ state (which has a gap and therefore no low energy quasiparticle excitations).  The reason that this question might be asked is  that there is at
present no detailed analysis indicating what the signature of a $d+id^\prime$ superconductor would be in a tricrystal
experiment.  Furthermore, as noted above, theoretical arguments have been made that the broken time-reversal symmetry $d+id^\prime$ state could play a role in
high T$_c$ superconductivity.  Clearly some analysis of the signature of $d+id^\prime$ superconductivity in phase-sensitive pairing symmetry experiments
would be useful in resolving these questions.

The above discussion indicates considerable current interest in, 
and motivation for, the
study of various types of broken time-reversal symmetry states.
The main objective of this article is therefore to study how broken
time-reversal symmetry superconducting states of $d + id'$ and $d+is$
symmetry, in both their bulk and surface forms, would manifest
themselves in phase-sensitive pairing-symmetry experiments of the tetracrystal type\protect\cite{tsu00}. (For reasons discussed in the following paragraph,
the focus will be on the tetracrystal geometry only, and no discussion will
be given of the tricrystal geometry, which is the geometry that has been
used in most of the
phase sensitive pairing symmetry experiments carried out to date\protect\cite{tsu00}.)
Some aspects related to the bulk $d + is$ state have already been
discussed \protect\cite{wal96}.  The details of the $d + id'$ case are
somewhat different, however. As a result, the $d + id^\prime$ and the
$d + is$ states give quite different results for an appropriately
chosen experimental geometry and these differences provide a method of
experimentally identifying a state of $d+id^\prime$ symmetry.

\begin{figure}[t!] % fig 1
\hspace{0 in}
\centerline{\epsfig{file=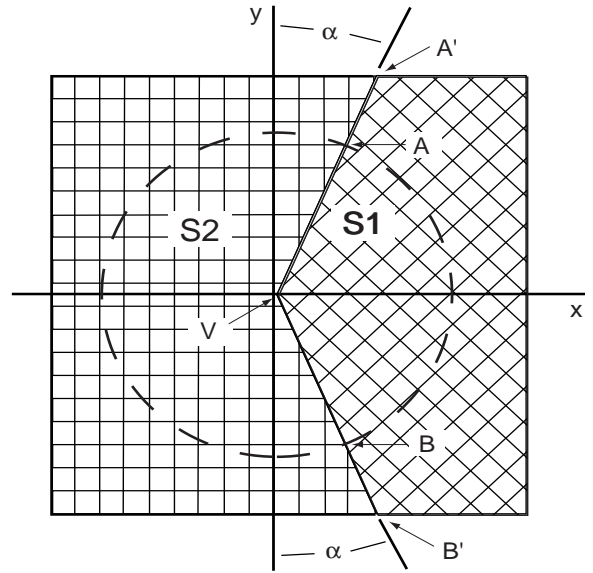,height=3in,width=3in}}
%\vspace{10pt}
\vspace{0 in}
\caption{The tetracrystal geometry\protect\cite{wal96,tsu97}. The
substrate is made from two tetragonal crystals of the same material
(e.g. SrTiO$_3$), S1 and S2, with their c-axes normal to the plane of
the paper, and their tetragonal a and b axes oriented at 45 degrees to
each other, as indicated by the cross ruling in the figure.  For
technical reasons, the configuration in the figure is easiest to
manufacture if the crystals S1 and S2 are each composed of two
separate crystals that are fused together along the line that is the
x-axis in the figure.  For this reason, this is called the
tetracrystal geometry.  A superconducting layer is deposited
epitaxially on the substrate, and the grain boundary A$^\prime$AVBB$^\prime$ forms a
Josephson junction.  The dashed circle is not part of the experimental
setup, but its arcs give the paths of a contour intergrals described
in the text.}
\label{fig1}
\end{figure}

The symmetric tetracrystal configuration proposed in Ref.\
\protect\onlinecite{wal96} and realized experimentally in Refs.\
\protect\onlinecite{tsu97,sch00} is shown
in Fig.\ \ref{fig1}.  This is called a symmetric geometry because the
x axis of the figure is an axis of reflection symmetry, and this
symmetry plays an essential role both in simplifying the analysis, and in
allowing an unambiguous interpretation of the experimental results.
The reason for choosing to analyze the tetracrystal geometry (rather than
the tricrystal geometry) in this article is that this geometry allows
a determination to be made of order parameter symmetry from the observed flux
quantization (say half integral) based on symmetry arguments only.  In the
tricrystal experiment, symmetry arguments can not be used because the
experimental arrangement has no useful symmetry elements (such as a plane
of reflection symmetry).  It is therefore necessary to appeal to a
phenomenological model for the dependence of the Josephson critical current on
the misorientation angles $\Theta_1$ and $\Theta_2$ of the two grains on either side of the grain boundary Josephson junction. For example, an appropriate
model accounting for the known experimental results on cuprate superconductors
would be
\begin{equation}
	J_c(\Theta_1, \Theta_2) = C(\Theta_1, \Theta_2)
		f_{sym}(\Theta_1, \Theta_2).
	\label{J_c}
\end{equation}
Here $C(\Theta_1,\Theta_2)$ is a positive factor taking account of the
changes in grain boundary microstructure with increasing misorientation angle
\protect\cite{hil99}; this effect is expected to dominate the overall magnitude
of the Josephson critical current, which is found experimentally to vary
roughly as
$\exp[-(\Theta_1 + \Theta_2)/\Theta_0]$ where $\Theta_0 \approx 5^\circ$ for hole-doped superconductors\protect\cite{hil98}
and $\Theta_0 \approx 2^\circ$ for electron-doped superconductors\protect\cite{sch99}. The
factor $f_{sym}$ can change sign and reflects the pairing symmetry of the
superconducting state. For the case of $d$-wave symmetry the simplest choices
are $f_{sym}=cos(2\Theta_1)cos(2\Theta_2)$\protect\cite{sig92} and the maximum disorder limit $f_{sym}=cos(2\Theta_1 + 2\Theta_2)
$\protect\cite{tsu94}. A systematic study of a variety of different
tricrystal geometries has established the reliability of this approach (see Ref.\ \protect\onlinecite{tsu00} for a review). In addition, the results of
the tricrystal experiments are in agreement with those obtained using the symmetric tetracrystal 
geometry\protect\cite{tsu00,tsu97,sch00} which, as noted above, does not refer
to a detailed model of the angular dependence of the Josephson critical 
current. 

The exploration of various broken time-reversal symmetry states carried out
in this article is more complex than testing for $d$-wave versus $s$-wave
symmetry, since both the $d+id^\prime$ and $d+is$ states involve linear
combinations of order parameters of two different symmetries. Determining
a quantitative formula for the angular dependence of the Josephson critical
current sufficiently  accurate for use with the tricrystal geometry is thus
even more difficult than for the purely $d$-wave and $s$-wave cases.  The
symmetric tetracrystal geometry presents no such difficulties, as accurate
predictions can be made on the basis of symmetry arguments only provided the
cuprate under study is tetragonal.    
\hspace{-14 pt}
In fact,
for the special case of $\alpha = 45^\circ$ in the symmetric tetracrystal
geometry, an additional symmetry comes into play
resulting in predictions
of remarkable simplicity (see Table\ \ref{table1} below).  Thus
the symmetric tetracrystal geometry, which has certain advantages relative
to the tricrystal geometry even in the cases where no broken time-reversal
symmetry is envisaged \protect\cite{tsu00,tsu97,sch00}, should be the geometry of choice for the investigation of differences between broken-time reversal symmetry states.

The nature of the spontaneous flux generated at the vertex V in Fig.\
\ref{fig1} can be determined by using Ginzburg-Landau theory, as
described in Refs. \protect\onlinecite{wal96,and87,sig95}.  The
superconducting order is characterized by an order parameter having
two components $\psi_d$ and $\psi_{d^\prime}$. (This is for the $d +
id^\prime$ case; for the $d + is$ case the two components are called
$\psi_d$ and $\psi_s$.) Also, the value of the $\lambda$'th component
($\lambda$ is $d$ or $d^\prime$) of the order parameter at the point J
(J is A or B) in the j'th superconductor (j is 1 or 2 for S1 or S2) is
denoted by $\psi_{\lambda J}^{(j)}$.
 
The free energy per unit length of the grain boundary Josephson
junction at point J is
\begin{equation}
	F_J = \sum_{\lambda,\lambda^\prime} C_{\lambda \lambda^\prime
	J} \left[\psi_{\lambda J}^{(1) *}\psi_{\lambda^\prime J}^{(2)}
	exp\left(i\frac{2\pi}{\Phi_0}\int_{J1}^{J2}{\bf A\cdot}d{\bf
	\ell}\right) + c.c.\right], \label{F_J}
\end{equation}
which is consistent with gauge invariance and time-reversal
symmetry. The line integral from point J1 to J2 in this equation is
across the grain boundary at point J from superconductor S1 to S2. The
fact that the x-axis is an axis of reflection symmetry imposes the
conditions $C_{\lambda \lambda A} = - C_{\lambda \lambda B}$ and
$C_{\lambda \lambda^\prime A} = C_{\lambda \lambda^\prime B}$ for
$\lambda \neq \lambda^\prime$.  (For the $d + is$ case the
corresponding relations are $C_{ddA} = - C_{ddB}, C_{ssA} = C_{ssB},
C_{sdA} = C_{sdB}$, and $C_{dsA} = - C_{dsB}$.) It is these symmetry
relations that allows a definitive interpretation of experiments
performed in the tetracrystal geometry to be made without recourse to
a detailed model of the angular dependence of the Josephson junction
current\protect\cite{wal96}.

The assumption of a $d+id^\prime$ state is implemented by taking the
order parameters to have the form
\begin{equation}
	\psi_{dJ}^{(j)} =  |\psi_{dJ}^{(j)}|e^{i\phi_{Jj}},\ \ 
	\psi_{d^\prime J}^{(j)} =  \sigma_{Jj} i|\psi_{dJ}^{(j)}|e^{i\phi_{Jj}}.
	\label{psi}
\end{equation}
Note the additional factor of $i$ (which characterises the
$d+id^\prime$ state) in the definition of $\psi_{d^\prime
J}^{(j)}$. Also, $\sigma_{Jj}$ can have the values $\pm 1$
corresponding to the two possible variants $d \pm id^\prime$, each of
which has the same free energy. For a bulk broken time-reversal
symmetry state in, say, superconductor S1, this state would be
expected to be characterized by either the order parameter
$d+id^\prime$ or $d-id^\prime$ throughout the entire crystal S1
because a change from one variant to the other would cost energy.
Thus we would have $\sigma_{A1} = \sigma_{B1}$, i.e. that same state
of superconductor S1 at both junctions.

On the other hand, it is also interesting to consider the case of a
superconductor that has pure $d$-wave symmetry in the bulk, but in
which the $d$-wave superconductivity is suppressed at the junction
interface with the simultaneous appearance of a $d^\prime$ order
parameter there, so that at the junction interface one has $d\pm
id^\prime$ superconductivity.  In this case one might have
$d+id^\prime$ surface superconductivity in superconductor S2 at
junction A and $d-id^\prime$ surface superconductivity in
superconductor S2 at junction B, so that $\sigma_{A2}$ and
$\sigma_{B2}$ are not necessarily equal.

Now put the above expressions for the order parameter into the
junction free energy of Eq.\ \ref{F_J}, and define a critical current
per unit length along the grain boundary, $I_A$, and a phase
$\theta_A$, characteristic of the grain boundary Josephson junction at
A by
\begin{equation}
	\frac{\Phi_0}{4\pi c} I_A e^{i\theta_A} \equiv g_{dd} 
	+ \sigma_{A1} \sigma_{A2} g_{d^\prime d^\prime}
	+i(\sigma_{A2} g_{d d^\prime} - \sigma_{A1} g_{d^\prime d}),
	\label{I_a}
\end{equation}
where 
\begin{equation}
	g_{\lambda \lambda^\prime} \equiv C_{\lambda \lambda^\prime A} 
	|\psi_{\lambda A}^{(1)}| |\psi_{\lambda^\prime A}^{(2)}|.
	\label{g}
\end{equation}
Similarly, for point B on the grain boundary, define $I_B$ and $\theta_B$ by
\begin{equation}
	\frac{\Phi_0}{4\pi c} I_B e^{i\theta_B} \equiv g_{dd} 
	+ \sigma_{B1} \sigma_{B2} g_{d^\prime d^\prime}
	-i(\sigma_{B2} g_{d d^\prime} - \sigma_{B1} g_{d^\prime d}).
	\label{I_b}
\end{equation}
The fact that the $x$ axis of Fig.\ \ref{fig1} is a line of reflection symmetry
has the important consequence that the same coefficients $g_{\lambda \lambda^\prime}$ that appear in Eq.\ \ref{I_a} also appear in Eq.\ \ref{I_b}.
The grain boundary junction free energies per unit length at points A and B can now be written
\begin{equation}
	F_A = \frac{\Phi_0}{2\pi c} I_A cos(\phi_A + \theta_A),
	\label{F_A}
\end{equation}
and
\begin{equation}
	F_B = \frac{\Phi_0}{2\pi c} I_B cos(\phi_B + \theta_B + \pi).
	\label{F_B}
\end{equation}
where
\begin{equation}
	\phi_J = \phi_{J2} - \phi_{J1} 
	+ \frac{2\pi}{\Phi_0}\int_{J1}^{J2}{\bf A\cdot}d{\bf \ell}.
	\label{phi}
\end{equation}

Note that the values of $\sigma_{Jj}$ in Eqs.\ \ref{I_a} and \ref{I_b}
have not yet been chosen.  These should be chosen so that the critical
currents $I_A$ and $I_B$ have their maximum possible magnitudes, thus
leading ultimately to the lowest possible free energy for the system.

A measurement of the spontaneous magnetic flux through the
grain-boundary Josephson junction of Fig.\ \ref{fig1} in a direction
normal to the page shows that the flux is concentrated in the
neighborhood of the vertex V, and falls exponentially to zero as one
proceeds along the grain boundary in the direction of either A or
B\protect\cite{tsu00}.  The characteristic length scale of this fall
off is the Josephson penetration depth.  It is assumed that the points
A and B are many Josephson penetration depths away from the vertex V
so that there is no flux through the grain boundary at points A and B,
nor is there any current flowing across the grain boundary at these
points.  The free energies $F_A$ and $F_B$ will therefore have their
minimum values at these points, leading to the results
\begin{equation} 
	\phi_A + \theta_A +\pi = 0,\ \ \phi_B + \theta_B = 0,\ \ (mod 2\pi).
	\label{phi_min}
\end{equation}
Next note that since the dashed circle is many penetration depths from
the vertex V, there are no currents flowing in the superconductor in
its vicinity.  It follows that
\begin{equation}
	\phi_{Aj} -\phi_{Bj} + \frac{2\pi}{\Phi_0}\int_{Bj}^{Aj} 
		{\bf A}\cdot d{\bf \ell} =0.
	\label{phi_sup}
\end{equation}
where the line integral follows an arc of the dashed circle in Fig.\
\ref{fig1}.  Finally, combining Eqs.\ \ref{phi}, \ref{phi_min}, and
\ref{phi_sup} leads to the result that the spontaneous flux through
the grain boundary junction at the vertex V is
\begin{equation}
	\Phi_{spont} = \left[n + \frac{1}{2} - \frac{\theta_B - \theta_A}{2\pi}\	\right]\Phi_0.
	\label{Phi_spont}
\end{equation}
For the sake of definiteness, choose $\theta_A$ and $\theta_B$ (which previously
have been defined mod $2\pi$) such that $0 < (\theta_B - \theta_A) \le 2\pi$.
Also, up to now the magnetic energy $\Phi^2/(2 L)$ has been assumed to be small in comparison with the free energies of Eqs.\ \ref{F_A} and \ref{F_B}, and has been left out of the problem. Since the magnetic free energy is a positive definite
function of the flux $\Phi$, the state (i.e. the value of $n$ in Eq.\ \ref{Phi_spont}) giving the lowest value of the total free energy is the one for which the flux is
\begin{equation}
	\Phi_{min} = \left[ \frac{1}{2} - \frac{\theta_B - \theta_A}{2\pi}\	\right]\Phi_0.
	\label{Phi_min}
\end{equation}
This minimum flux is such that $-\Phi_0/2 < \Phi_{min} \le \Phi_0/2$.  Thus
the state corresponding to $n=0$ in Eq.\ \ref{Phi_spont} is the ground state,
and the states corresponding to n a nonzero integer are possible metastable states.

In discussing the bulk $d\pm id^\prime$ states, the requirements
$\sigma_{Aj} = \sigma_{Bj}$ (see above) are inserted into Eqs.\
\ref{I_a} and \ref{I_b}, yielding $\theta_B = -\theta_A$. An
interesting limit is that in which the $d$ component of the order
parameter is assumed to be much greater than the $d^\prime$ component.
An examination of Eqs.\ \ref{I_a} and \ref{I_b} in this case shows
that, so long as $\alpha$ is not too close to $45^\circ$, the
term in $g_{dd}$ is expected to dominate on the right hand side, and
hence $|\theta_A| = |\theta_B|$ will be much less than $\pi$. This
results in an approximately half integral flux quantum effect, which
is not very different from what is expected for a pure $d$ state. The
departure of the spontaneously generated flux from $\Phi_0/2$ will
grow as the magnitude of $\psi_{d^\prime}$ grows relative to that of
$\psi_d$.  Thus, to detect broken time-reversal symmetry with
$|\psi_{d^\prime}| \ll |\psi_d|$ requires a high precision measurement
of the spontaneous flux, because the signature of the presence of a
small $d^\prime$ component of the order parameter is a small departure
of the spontaneous flux from $\Phi_0/2$. For $\alpha = 45^\circ$ however,
an additional symmetry of the Josephson junctions requires that
$g_{dd} = g_{d^\prime d^\prime} = 0$. This yields $\theta_A =
-\theta_B = \pm \pi/2$ and a spontaneous flux $\Phi = n \Phi_0$, which
is precisely the same result as would be obtained for a classical $s$-wave 
superconductor.  The result that the broken time-reversal
symmetry $d+id^\prime$ state has $\Phi = n \Phi_0$ as its flux quantization
condition for this geometry is quite remarkable
since broken time-reversal symmetry states
are generally believed to have the flux quantization condition 
$\Phi = (n + f) \Phi_0$, where the fraction $f$ 
satisfies $-1/2 < f \le 1/2$ and
is neither precisely 0 nor precisely 1/2.  In summary note that, for the
$d + id^\prime$ state, the geometries with $\alpha \sim 22^\circ$ and
$\alpha = 45^\circ$ lead to completely different flux
quantization conditions, and that these different conditions 
give a way of experimentally identifying
a $d+id^\prime$ state.

The introduction to this article cites several theoretical studies
that suggest a bulk $d$-wave superconductor will tend to exhibit a
suppressed $d$-wave order parameter at a [110] surface, possibly
accompanied by the presence of a $d'$ or $s$ component at this surface,
and with these components
phased in such a way that the overall state at the surface has
$d+id^\prime$ or $d+is$ symmetry. An appropriate geometry in which
to study such states is to take the angle $\alpha$ in Fig.\ \ref{fig1}
to be 45 degrees. There is then the possibility of a surface $d+id^\prime$
or surface $d+is$ state in the superconductor S2 (but not S1) along the grain
boundaries A$^\prime$AV and VBB$^\prime$.  There are two inequivalent possibilities for the
surface states of the two grain boundary segments: A$^\prime$AV can be $d\pm id^\prime$
while VBB$^\prime$ is also $d\pm id^\prime$, or A$^\prime$AV can be $d\pm id^\prime$
while VBB$^\prime$ is  $d\mp id^\prime$ (here take only the upper signs, or only
the lower signs).  There is an interaction energy at the
vertex V which should give these two inequivalent possibilities different free energies,
but this interaction energy will be much smaller than the Josephson energy of
the entire grain boundary and therefore the assumption will be made below that
both possibilities can occur for  the same sample.

From the work already done above, it follows that the case of purely $d$-wave symmetry in the bulk, with $d\pm
id^\prime$ surface superconductivity at the [110] surface of
superconductor S2 for the case of $\alpha = 45^\circ$ also yields a flux
quantization condition $\Phi = n \Phi_0$ for the case $\sigma_{A2} = \sigma_{B2}$, but with the new feature that there is the
possibility of a states of the same sample having the flux quantization
condition $\Phi = (n + \frac{1}{2})\Phi_0$ for
$\sigma_{A2} = - \sigma_{B2}$.  Of all of these states, the ground state is
the one with zero flux.  Again, it is remarkable that the broken time-reversal
symmetry surface states for this geometry lead to flux quantization conditions
usually thought to be characteristic of states without broken time reversal
symmetry.

The analysis of the $d + is$ case is similar, except that the defining
equations for the critical currents and characteristic phases, Eqs. \ref{I_a} and
\ref{I_b}, are replaced by
\begin{equation} \frac{\Phi_0}{4\pi
	c} I_{A} e^{i\theta_{A}} \equiv g_{dd} - \sigma_{A1}
	\sigma_{A2} g_{ss} +i(-\sigma_{A1} g_{sd} + \sigma_{A2}
	g_{ds}), \label{I_A}
\end{equation}
and
\begin{equation}
	\frac{\Phi_0}{4\pi c} I_{B} e^{i\theta_{B}} \equiv g_{dd} 
	+ \sigma_{B1} \sigma_{B2} g_{ss}
	+i(\sigma_{B1} g_{sd} + \sigma_{B2} g_{ds}).
	\label{I_B}
\end{equation}
Arguments similar to those given above lead to the result that the
spontaneous flux at the vertex V in Fig.\ \ref{fig1} is again given by
Eq.\ \ref{Phi_spont}.

Now consider the limit of the bulk $d\pm is$ case in which the
magnitude of the $d$ component of the order parameter is much greater
than that of the $s$ component.  An examination of Eqs.\ \ref{I_A} and
\ref{I_B} in this case shows that, so long as $\alpha$ is not too
close to $45^\circ$, the term in $g_{dd}$ is expected to dominate
the right hand side, and hence $|\theta_A|$ and $|\theta_B|$ will be
much less than $\pi$. This results in an approximately half integral
flux quantum effect, which is not very different from what is expected
for a pure $d$ state, nor from the $d+id^\prime$ case for the
corresponding values of $\alpha$.  Now note that for $\alpha = 45^\circ$,
the additional symmetry of the Josephson junctions requires that
$g_{dd} = g_{sd} = 0$. (Also, $|g_{ss}| \ll |g_{ss}|$.) This yields
$\theta_A \simeq \theta_B \simeq \pi/2$ or $\theta_A \simeq \theta_B
\simeq -\pi/2$. Hence for $\alpha = 45^\circ$ also the approximate half
integral flux quantum effect is preserved, which is very different
from what was predicted above for the bulk $d + id^\prime$ case.
Thus, the symmetric tetracrystal geometry for $\alpha = 45^\circ$ can in
principle distinguish between the bulk $d + id^\prime$ and $d + is$
states.

The case of purely $d$-wave symmetry in the bulk, with $d\pm is$
surface superconductivity at the [110] surface of superconductor S2
for the case of $\alpha = 45^\circ$ yields the flux quantization conditions
of $\Phi = n \Phi_0$ for $\sigma_{A2} = \sigma_{B2}$ and
$\Phi = (n + \frac{1}{2}) \Phi_0$ for $\sigma_{A2} = -\sigma_{B2}$.  These
and other results are summarized in Table\ \ref{table1}.

The grain boundaries that occur in Fig.\ \ref{fig1} for $\alpha =
45^\circ$ are called $45^\circ$ [001] tilt grain boundaries. Scanning
SQUID microscope imaging of such grain boundaries in
YBa$_2$Cu$_3$O$_{6+x}$ reveals very weak spontaneously generated
delocalized flux along these grain boundaries\protect\cite{man96}.
This is thought to be due to either grain-boundary

\begin{table} %[t!]
\caption{The flux quantization conditions for the tetracrystal geometry
of Fig.\ \ref{fig1} with the angle $\alpha = 45^\circ$ for a number
of different broken time-reversal symmetry states. For the surface case,
the two states given represent the surface states in superconductor S2
on grain boundaries A and B, respectively, of Fig.\ \ref{fig1}. Also,
$f$ satisfies $-\frac{1}{2} < f < \frac{1}{2}$, but is not zero; 
although $f$ can not be precisely $\frac{1}{2}$, it is expected to
be  close to $\frac{1}{2}$ if the time-reversal symmetry breaking is weak.}
\begin{tabular}{ll} 
state ($\alpha = 45^\circ$ in all cases)	&	flux quantization \\ \hline
bulk 	$d + id^\prime$				&	$n\Phi_0$ 	\\
bulk	$d + is$					&	$(n+f)\Phi_0$	\\
surface $d + id^\prime$,\  $d + id^\prime$ & $n\Phi_0$ \\
surface $d + id^\prime$,\  $d - id^\prime$ & $(n+\frac{1}{2})\Phi_0$ \\
surface $d + is$,\  $d + is$ 			& $(n+\frac{1}{2})\Phi_0$ \\
surface $d + is$,\  $d - is$ 			& $n\Phi_0$ \\
\end{tabular}
\label{table1}
\end{table}
\hspace{-14 pt}
faceting\protect\cite{man96}, or the existence of orthorhombic twins
\protect\cite{wal96b} in the YBa$_2$Cu$_3$O$_{6+x}$, or both. To avoid
this effect (which can be detected using scanning SQUID microscopy as
just mentioned) the material studied should be tetragonal and not
orthorhombic (thus avoiding twinning) and the grain boundaries should
be as straight as possible.  In the case of $d+is$ and $d+id^\prime$
states, the 
additional component of the order parameter (in addition
to the $d$-wave component) provides an additional coupling which, if
large enough, will overcome the effects giving rise to the spontaneous
delocalized grain boundary flux (which results from the fact that
usual $d$-wave Josephson coupling is zero for an ideally flat
$45^\circ$ [001] tilt grain boundary).

This article has considered
measurements of the pairing symmetry made using the tetracrystal
geometry. For the bulk $d+id^\prime$ and $d+is$ states, and for
the value $\alpha \sim 22^\circ$ (see Fig.\ \ref{fig1}) advocated in Refs.\
\protect\onlinecite{wal96} and \protect\onlinecite{tsu97}, the 
flux quantization condition is 
$\Phi = (n+f)\Phi_0$, where $f$ is a fraction satisfying
$-1/2 < f < 1/2$; also $f \ne 0$.  If the magnitude 
of the symmetry breaking is small, $f$ is expected
to be close to $1/2$, and the measurements must be relatively 
high precision to detect the small differences of the spontaneous flux
from the value of $\Phi_0/2$.  However, striking differences
between the bulk $d+id^\prime$ and $d+is$ states occur in the highest
symmetry case where
the angle $\alpha = 45^\circ$. In this case, the $d + id^\prime$
state should exhibit a spontaneous flux of zero while the $d+is$ state
should exhibit a spontaneous flux of approximately $\Phi_0/2$. This should
provide a way of experimentally identifying a state with $d+id^\prime$ 
symmetry. For
bulk superconductivity that is purely $d$-wave, and surface
superconductivity of the type $d+id^\prime$ or $d+is$ at junctions A
and B in superconductor S2 of Fig.\ \ref{fig1} with $\alpha = 45^\circ$ ,
the flux quantization may be either of the type $\Phi = n\Phi_0$
or of the type $\Phi = (n + \frac{1}{2})\Phi_0$, depending on whether
the surface states on the two differently oriented grain boundaries in
Fig.\ \ref{fig1} are the same, or are time-reversed conjugates. 
Interestingly, in this
highest symmetry geometry, flux quantization conditions usually
expected for states without broken time-reversal symmetry [i.e.
$\Phi = \Phi_0$ or $\Phi = (n + \frac{1}{2})\Phi_0$] are also found for
several broken time-reversal symmetry states.

I would like to thank J. Y. T. Wei and C. C. Tsuei for the discussions
that led to this article, C. C. Tsuei and L. Taillefer for useful comments
on the manuscript, and the Natural Sciences and Engineering
Research Council for support.

%\begin{figure}
%\caption{A model}
%\label{fig1}
%\end{figure}


\begin{references}

\bibitem{tsu00} C. C. Tsuei and J. R. Kirtley, ``Pairing symmetry in
  cuprate superconductors,'' to appear in Rev. Mod. Phys.

\bibitem{bal98} A. V. Balatsky, Phys. Rev. Lett. {\bf 80}, 1972 (1998).

\bibitem{rok93} D. S. Rokhsar, Phys. Rev. Lett. {\bf 70}, 493 (1993).

\bibitem{lau98}  R. B. Laughlin, Phys. Rev. Lett. {\bf 80}, 5188 (1998).

\bibitem{mov98} R. Movshovich, M. A. Hubbard, M. B. Salamon, A. V. Balatsky, 
R. Yoshizaki, J. L. Sarrao, and M. Jaime, Phys. Rev. Lett. {\bf 80}, 1968 (1998).

\bibitem{kri98} K. Krishana, N. P. Ong, q. Li, G. D. Gu, and N. Koshizuka,
Science {\bf 277}, 83 (1998).

\bibitem{li93} Q.P. Li, B. E. C. Koltenbah, and R. Joynt, Phys. Rev. B {\bf 48},
437 (1993).

\bibitem{mat95} M. Matsumoto and H. Shiba, J. Phys. Soc. Japan {\bf
64}, 3384, 4867 (1995).

\bibitem{sal98} M. I. Salkola and J. R. Schrieffer, Phys. Rev. {\bf
58}, R5952 (1998).

\bibitem{kas95} K. Kashiwaya, Y. Tanaka, M. Koyanagi, H. Takashima and
K. Kajimura, J. Phys. Chem. Solids {\bf 56}, 1721 (1995).

\bibitem{fog97} M. F\"ogelstrom, D. Rainer, and J. A. Sauls,
  Phys. Rev. Lett {\bf 80}, 4763 (1998).

\bibitem{zhu98} J.-X. Zhu and C. S. Ting, Phys. Rev. B {\bf 57}, 3038
(1998).

\bibitem{zhu99} J.-X. Zhu, B. Friedman and C. S. Ting, Phys. Rev. B {\bf 59},
	3353 (1999).
  
\bibitem{cov97} M. Covington, M. Aprili, E Paraoanu, L. H. Greene, F.
  Xu, J. Zhu, and C. A. Mirkin, Phys. Rev. Lett. {\bf 79}, 277 (1977).

\bibitem{tsu00b} C. C. Tsuei and J. R. Kirtley, cond-mat/0002341.

\bibitem{tai00} L. Taillefer, reported at the M2S-HTSC-VI Conference on
Superconductivity in Houston, February 2000.

\bibitem{hus00} N. E. Hussey, K. Behnia, H. Takagi, C. Urano, S. Adachi, and S. Tajima, cond-mat/0004094

\bibitem{wal96} M. B. Walker and J. Luettmer-Strathmann, Phys. Rev. B
{\bf 54}, 588 (1996).
  
\bibitem{tsu97} C. C. Tsuei, J. R. Kirtley, Z. F. Ren, J. H. Wang, H.
  Raffy, and Z. Z. Li, Nature {\bf 387}, 481 (1997).

\bibitem{sch00} R. R. Schulz, B. Chesca, B. Goetz, C. W. Schneider, A. Schmehl, H. Bielefeldt, H. Hilgenkamp, J. Mannhart and C. C. Tsuei, App. Phys. Lett. {\bf 76}, 912 (2000).

\bibitem{hil99} H. Hilgenkamp, C. W. Schneider, B. Goetz, R. R. Schulz,
A. Schmehl, H. Bielefeldt and J. Mannhart, Supercond. Sci. Technol. {\bf 12}, 1043 (1999).

\bibitem{hil98} H. Hilgenkamp and J. Mannhart, App. Phys. Lett. {\bf 73}, 265 (1998).

\bibitem{sch99} U. Schoop, S. Kleefisch, S. Meyer, A. Marx, L. Alff and R. Gross,  IEEE Trans. on Appl. Supercond. {\bf 9}, 3409 (1999).

\bibitem{sig92} M. Sigrist and T. M. Rice, J. Phys. Soc. Japan {\bf 61}, 4283 (1992).

\bibitem{tsu94} C. C. Tsuei, J. R. Kirtley, C. C. Chi, L. S. Yu-Jahnes, A. Gupta, T. Shaw, J. Z. Sun and M. B. Ketchen, Phys. Rev. Lett. {\bf 73},
593 (1994).

\bibitem{and87} A. F. Andreev, JETP Lett. {\bf 46}, 584 (1987).

\bibitem{sig95} M. Sigrist and T. M. Rice, Rev. Mod. Phys. {\bf 65},
503 (1995).

\bibitem{man96} J. Mannhart, H. Hilgenkamp, B. Mayer, Ch. Gerber,
J. R. Kirtley, K. A. Moler, and M. Sigrist, Phys. Rev. Lett. {\bf 77},
2782, (1996).

\bibitem{wal96b} M. B. Walker, Phys. Rev. B {\bf 54}, 13269, (1996).
 
\end{references}
\end{document}